\documentclass[sigconf,nonacm]{acmart}

\AtBeginDocument{%
  }

\setcopyright{none}
\renewcommand\footnotetextcopyrightpermission[1]{}
\acmConference[ ]{}{}{}  

\usepackage{url}
\usepackage{framed}

\newtheorem{definition}{Def.}

\usepackage{titlesec}
\titlespacing*{\section}{0pt}{0.5em}{0.1em}
\titlespacing*{\subsection} {0pt}{0.5em}{0.1em}
\titlespacing*{\subsubsection} {0pt}{0.1em}{0.1em}
\begin{document}

\title{HotBugs.jar: A Benchmark of Hot Fixes for Time-Critical Bugs}

% \affiliation{%

% \affiliation{%

\author{Carol Hanna, Federica Sarro, Mark Harman, Justyna Petke}
\affiliation{%
  \institution{University College London}
  \city{London}
  \country{United Kingdom}
}

% \affiliation{%

% \affiliation{%

% \affiliation{%

% \affiliation{%

% \affiliation{%

\renewcommand{\shortauthors}{Hanna et al.}

\begin{abstract}
Hot fixes are urgent, unplanned changes deployed to production systems to address time-critical issues.
Despite their importance, no existing evaluation benchmark focuses specifically on hot fixes.
We present HotBugs.jar, the first dataset dedicated to real-world hot fixes.
From an initial mining of 10 active Apache projects totaling over 190K commits and 150K issue reports, we identified 746 software patches that met our hot-fix criteria.
After manual evaluation, 679 were confirmed as genuine hot fixes, of which 110 are reproducible using a test suite.
Building upon the Bugs.jar framework, HotBugs.jar integrates these 110 reproducible cases and makes available all 679 manually validated hot fixes, each enriched with comprehensive metadata to support future research.
Each hot fix was systematically identified using Jira issue data, validated by independent reviewers, and packaged in a reproducible format with buggy and fixed versions, test suites, and metadata.
HotBugs.jar has already been adopted as the official challenge dataset for the Search-Based Software Engineering (SBSE) Conference Challenge Track, demonstrating its immediate impact.
This benchmark enables the study and evaluation of tools for rapid debugging, automated repair, and production-grade resilience in modern software systems to drive research in this essential area forward.
\end{abstract}

%%z



\maketitle

\section{Introduction}
Modern software systems are increasingly complex, distributed, and continuously deployed, making failures inevitable even with rigorous testing and quality assurance practices. When critical issues arise in production, development teams must respond quickly by issuing hot fixes. Hot fixes are unplanned changes that address specific urgent, time-sensitive bugs in production. Hot fixes play a crucial role in maintaining the reliability, performance, and security of software systems, particularly in large-scale platforms used by thousands of users or mission-critical environments. However, despite their importance, hot fixes have received limited attention in empirical software engineering research. Most existing studies focus on general bug fixing or maintenance practices, leaving a gap in understanding the unique characteristics, urgency, and challenges associated with hot fix scenarios.

Benchmarking has proven to be a cornerstone for advancing research in areas such as automated program repair, fault localization, and bug detection. Well-established datasets like Bugs.jar, Defects4J, and BEARS have enabled the community to develop and evaluate techniques in a reproducible and standardized manner. Yet, these benchmarks primarily contain bugs identified through planned development cycles or continuous integration pipelines. They lack the real-world urgency and contextual constraints of hot fixes, making them less suitable for studying time-critical maintenance tasks or evaluating tools designed for rapid response in production environments.

%% Mark added this to further motivate and increase adoption
The hot fixing benchmark represents new evaluation challenges for both test generation and for repair techniques.
Tests (such as just-in-time tests\cite{harman:harden}) 
for these `hot' bugs must be generated quickly, 
and any repair must be deployed quickly, as well as being effective at fixing the issue.
%% Mark also added this to further motivate and increase adoption
The benchmark also creates new opportunities for evaluation: 
For hot fixing, an initial partial fix that merely {\em ameliorates} the failure symptoms can be useful due to urgency.
Such a symptom-relieving fix can be deployed as an {\em initial}  hot fix 
to reduce the impact of the failure, while a longer term, more durable and complete, fix is found. 
In this way,  a hot fix benchmark suite not only challenges testing and repair research, 
but also provides a way to evaluate new potential use cases such as failure amelioration.

To bridge this gap, we present HotBugs.jar~\cite{hotbugs}, a novel benchmark specifically focused on hot fixes. HotBugs.jar is built as an extension of Bugs.jar~\cite{bugsdotjar} and provides a curated collection of real-world hot fixes drawn from mature, actively maintained open-source projects within the Apache ecosystem, such as Kafka, Flink, Hadoop, and Ambari. Each hot fix is systematically identified through structured issue-tracking data from Jira, filtered using a rigorous definition that captures the essential attributes of a hot fix: specificity, unplanned nature, time-criticality, and deployment in production. Furthermore, each candidate fix is manually validated by independent evaluators to ensure accuracy and reliability.
Even in its initial release, HotBugs.jar has already seen use by the broader software engineering research community.
Version 1 of the benchmark was selected as the challenge dataset for the Search-Based Software Engineering (SBSE) Conference Challenge Track, where it served as the foundation for evaluating innovative search-based techniques for time-critical bug fixing. 
This early adoption demonstrates the practical relevance and immediate applicability of HotBugs.jar.
This further validates its design and establishes it as a trusted resource for both researchers and practitioners.

\section{HotBugs.jar Construction}
In this section, we outline the methodology for collecting the hot fixes and constructing the HotBugs.jar benchmark. Figure~\ref{fig:hotfixarch}  visualizes this pipeline.

\subsection{Project Sourcing}
\label{projectSelection}
% Criteria:
% - Open source on github 
% - Jira issue tracker
% - Over 10K commits.
HotBugs.jar is based on  10 subject systems, all under the Apache ecosystem. The projects included are Kafka~\cite{kafka}, Flink~\cite{flink}, Hadoop~\cite{hadoop}, Solr~\cite{solr}, Hbase~\cite{hbase}, Jackrabbit~\cite{jackrabbit}, Karaf~\cite{karaf}, Nifi~\cite{nifi}, Ambari~\cite{ambari}, and Calcite~\cite{calcite}.
These projects were selected based on maturity and active maintenance.

\begin{figure*}[t]
\centering
\begin{minipage}{0.38\textwidth}\center
\includegraphics[width=\linewidth]{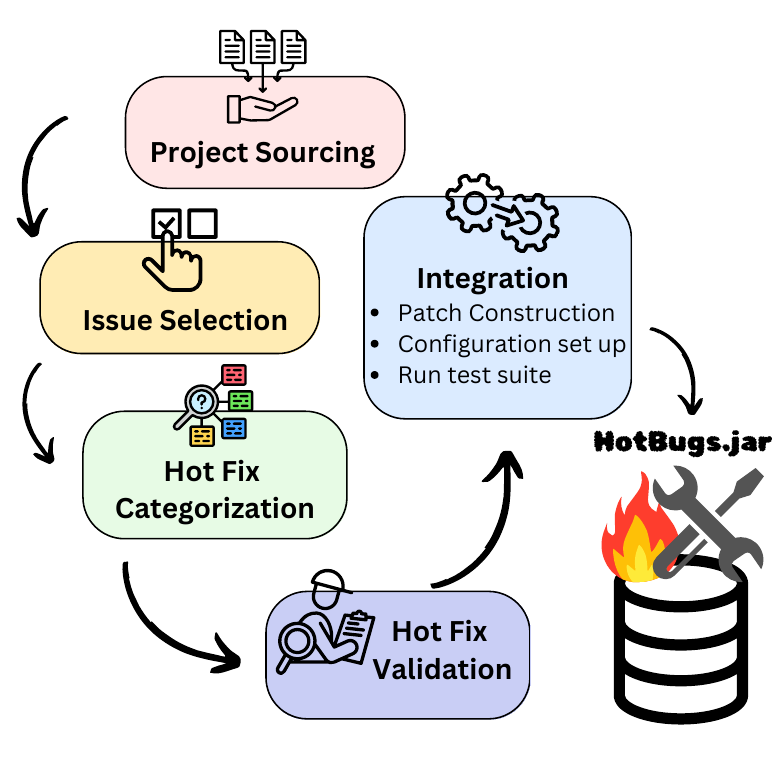}
  \caption{HotBugs.jar Construction}
  \label{fig:hotfixarch}
\end{minipage}
\begin{minipage}{0.60\textwidth}
  \captionof{table}{Specifications of subject programs (at date of collection) covering 10 large, actively maintained Apache projects totaling more than 190K commits and 150K issues.} 
  \centering
  \label{tab:subjectPrograms}
  \begin{tabular}{llrrrrrrrr}
    \toprule 
    Software & Date & Rels. & Commits & Issues & Blocker & Major & Critical \\
    \midrule
      Flink & 16/5/24 & 216 & 35295 & 22668 & 1035 & 3151 & 1156 \\ 
     \midrule
      Kafka & 04/3/24 & 213 & 12432 & 7708 & 599 & 2029 & 278 \\
    \midrule
     Hadoop& 19/6/24 & 342 & 27330 & 24928 & 818 & 5915 & 987\\
     \midrule
     Solr& 22/6/24 & 390 & 37511 & 24952 & 300 & 3732 & 188\\
     \midrule
     Hbase&24/7/24 & 238 & 20226 & 18768 & 508 & 4320 & 757\\
    %  Cassandra& 19/06/25& & & & & & & &\\
    % \midrule
    %  Fineract& 28/06/24& & & & & & & &\\
    % \midrule
    %  Hive& 13/06/25& & & & & & & &\\
    % \midrule
     % Ignite& 15/06/25 & 117 & 29420 & 9576 & 201 & 2410 & 399\\
    \midrule
     Jackrabbit& 17/6/24 & 264 & 9291 & 6175 & 26 & 825 & 48\\
    \midrule
    %  Jackrabbit-Oak& 13/06/25& & & & & & & &\\
    % \midrule
     Karaf& 13/6/24 & 99 & 9534 & 6755 & 33 & 1096 & 66\\
    \midrule
    Nifi & 28/6/25 & 134 & 10825 & 10260 & 208 & 2030 &  251 \\
    \midrule
    Ambari &  20/7/25 & 85 & 25069 & 24456 & 1304 & 9312 & 3703 \\
    \midrule
    Calcite & 28/6/25 &  90 & 6208 & 3915 &  35 & 1726 &  44\\
    \midrule
    \textbf{Total} &  &  \textbf{2071} & \textbf{193721} & \textbf{150585} &   \textbf{4866}&  \textbf{34136} & \textbf{7478} \\
    %  Spark& 26/06/24& & & & & & & &\\
    % \midrule
     % Groovy& 15/06/24& & & & & & & &\\
     \bottomrule

  \end{tabular}
\end{minipage}
\end{figure*}

The specifications of each project are outlined in Table~\ref{tab:subjectPrograms}.
For each project, we report the collection date, number of releases, total commits, and total Jira issues, as well as the breakdown of issues by priority level: Blocker, Major, and Critical.
The data show considerable variation across projects in terms of scale, with Ambari having the highest number of Jira issues and releases, while Calcite represents a smaller, yet still actively maintained, system.
This diversity ensures that the benchmark captures a wide range of real-world hot fix scenarios.

We followed specific criteria for inclusion to ensure that the hot fixes in our benchmark are representative of real-world hot fixing scenarios.
We look at open-source software on Github~\cite{github}.
Open source software is production code with real users being worked on by community-led maintainers.
Thus, well-maintained projects on GitHub tend to follow software engineering best practices.
From here, we only select well-maintained projects where the last commit is within the week of the project data getting collected.
We also focus on mature repositories where the number of commits exceeds 9K and repositories where the number of releases exceeds 100.

Finally, we only select projects that use the Jira issue tracking system~\cite{jira}.
Jira is one of the most popular systems for issue tracking both in open-source and commercial projects.
Issues on Jira are also very structured in format (e.g. fields for priority and severity labels, affected versions, creation and resolution dates, etc.).
This allows us to systematically extract all the relevant issue information in a uniform manner across different projects and ecosystems.

\subsection{Issue Collection}
%As per def:
%- unplanned: right after release
%- specific: jira issue
%- time criticality: priority tags + time to fix
%- issue: we look at bugs
%- deployed system: in deployed systems on github

We collect relevant issues according to Definition~\ref{def:hotfix} of hot fix.

\begin{framed}
\begin{definition}
\label{def:hotfix}
\noindent A \textbf{hot fix} is an unplanned improvement to a specific time-critical issue deployed to a software system in production.
\end{definition}
\end{framed}

Given this definition, we outline below how we consider all five factors that make up a hot fix:
\begin{enumerate}
    \item \textbf{Specific}: Bound within a single jira report.
    \item \textbf{Improvement}: issue label.
    \item \textbf{Unplanned}: created right after a release.
    \item \textbf{Time-critical}: priority labels and time to fix.
    \item \textbf{Deployed}: hot fixes in production.
\end{enumerate}

First, a hot fix is specific.
Thus, we only we only consider specific issues raised as a self-contained report on the Jira issue tracking system.
Additionally, since a hot fix mitigates a software issue, we only consider Jira reports tagged as bugs.

A hot fix refers to a required unplanned change.
To account for this, we check Jira issue created within a small delta of time from a release.
We specifically keep this delta small and look at issues created on the same day as a release.
In our data collection, this delta is equal to 24 hours.
Bugs found right after a release are likely unplanned changes.

Next, we consider time-criticality through two different lenses: syntactic and temporal.
Syntactically, we look at how the Jira issue is tagged in terms of priority.
We only consider bugs tagged as 'Blocker', 'Major', or 'Critical'.
These priority labels are the highest default priorities that the Jira platform offers.
Temporally, we look at the time difference between the creation of the Jira issue and its resolution.
These dates are provided as fields on the Jira platform.
Bugs that are reported and resolved within a short period of time are more likely to be time-critical issues that need hot fixing.
Thus, we only consider bugs created and resolved on the same day.

Finally, we only collect hot fixes from systems in production.
As explained in Section~\ref{projectSelection}, we only look at mature and currently well-maintained publically available GitHub projects.
These are projects are in production used and developed by the community.
To sum up, each Jira issue considered is tagged a \textit{Bug}, has a priority of Blocker/Major/Critical, is created on the same day of a release, and is created and resolved within the same day.

\subsection{Hot Fix Categorization}
We assess the bugs that fit the criteria outlined by the definition manually.
For each bug, we first categorize the bug type according to the bug taxonomy outlined by Catolino et al.~\cite{Catolino2019}.
According to this taxonomy, for each bug, we check whether it is an issue to do with configuration, database, GUI, network, performance, permission/depreciation, security, a test-related issue, or a program anomaly.
The details of this categorization are presented in Table~\ref{tab:hotFixCategories}.
The most common category across all projects is program anomalies (375), followed by test code issues (73) and configuration errors (96). Certain categories, such as GUI-related bugs and network issues, appear less frequently, indicating that hot fixes often target core functionality rather than peripheral concerns. 
For example, Ambari shows a broad distribution of hot fix types, with a notably high count of GUI-related hot fixes (47), while other projects such as Kafka and HBase primarily focus on program anomalies and high-priority backend issues.
\begin{table*}
  \caption{Hot Fix Categorization. Hot fixes are overwhelmingly programming-related, emphasizing backend stability over user-facing issues.} 
  \centering
  \label{tab:hotFixCategories}
  \begin{tabular}{lrrrrrrrrrrrr}
    \toprule 
    & \multicolumn{10}{c}{Hot Fix Categories} & \multicolumn{2}{c}{User-facing Hot Fix} \\
    \midrule 
    Software &
    Config. &
    Database &
    GUI &
    Network &
    Perf. &
    Permission/ &
    Security &
    Prog. &
    Test &
    Other &
    Yes &
    No
    \\
     &
     &
     &
     &
     &
     &
    Deprication &
     &
    anomaly &
    code &
     &
     &
    
    \\
    \midrule
     Flink & 4 & 0 & 0 & 1 & 0 & 4 & 0 & 20 & 3 & 2 & 7 & 27 \\
    \midrule    
     Kafka & 1 & 0 & 0 & 1 & 0 & 2 & 1 & 7 & 1 & 0 & 3 & 10 \\
    \midrule
     Hadoop  & 18 & 0 & 0 & 2 & 0 & 1 & 2 & 42 & 23 & 7& 42 & 53 \\
     \midrule
     Solr  & 11 & 1 & 0 & 0 & 5 & 0 & 2 & 33 & 11 & 3 & 19 & 47\\
     \midrule
     Hbase  & 3 & 0 & 0 & 0 & 0  & 0  & 0  & 15 & 3 & 1 & 7 & 15\\
     \midrule
     Jackrabbit & 1 & 2 & 0 & 0 & 1 & 3 & 2 & 15 & 3 & 0 & 2 & 25 \\
    \midrule
     Karaf & 8 & 0 & 0  & 0 & 0 & 0 & 2 & 8 & 1 & 1& 0 & 20 \\
    \midrule
    Calcite & 2 & 0 & 0 & 0 & 0 & 1& 0 & 8 & 2 & 0 &  4 & 9 \\
    \midrule
    Nifi & 10 & 0  & 3 & 2 & 5 & 6 & 0  & 29 & 8 & 1 &  15 & 49 \\
    \midrule
    Ambari & 38 & 4 & 47 & 2 & 3 & 9 & 5 & 198 & 18 & 1&  22 & 303 \\
    \bottomrule
    \textbf{Total}  & \textbf{96} & \textbf{7} & \textbf{50} & \textbf{8} & \textbf{14} & \textbf{26} & \textbf{14} & \textbf{375} & \textbf{73} & \textbf{16} &  \textbf{121} &  \textbf{558} \\
    \bottomrule
  \end{tabular}
\end{table*}
We also assess whether the issue is an internal or external hot fix.
Internal hot fixes are ones that target critical blocking issues within the system that affect the development team.
Such `internality' is often related to important testing issues or issues that block developers from resuming their work.
External hot fixes are ones that are user-facing.
For external hot fixes, we check whether the report specifically includes this.
The results of this assessment are presented in Table~\ref{tab:hotFixCategories}.

% \begin{table}
%   \caption{User-facing vs Internal Hot Fixing.} 
%   \centering
%   \label{tab:hotFixInternalExternal}
%   \begin{tabular}{lrrrrrrrrr}
%     \toprule 
%     Software  & \multicolumn{2}{c}{Hot Fix Type}\\
%      &  Internal & User-facing\\
%     \midrule
%      Flink   
%     \midrule    
%      Kafka
%     \midrule
%      Hadoop
%      \midrule
%      Solr
%      \midrule
%      Hbase
%      \midrule
%     % Cassandra& & & & & &\\
%     % \midrule
%     %  Fineract& & & & & &\\
%     % \midrule
%     %  Hive& & & & & &\\
%     % \midrule
%      Jackrabbi
%     \midrule
%     %  Jackrabbit-Oak& & & & & & \\
%     % \midrule
%      Karaf
%     \midrule
%     Calcite
%     \midrule
%     Nifi
%     \midrule
%     Ambari
%      % Spark & & & & \\
%      % Groovy & & & & \\
%     \bottomrule
%     \textbf{Total}
%     \bottomrule
%   \end{tabular}
% \end{table}

\subsection{Hot Fix Validation}
We validate the collected hot fixes by manually assessing them.
We check the Jira issue description, comments, and even the source code when necessary to understand the reported issue.
Given this information, we assess whether the issue does fit in with the hot fix definition and is indeed a true hot fix.
For each, we give a short description of why we flag it as a hot fix or not.

The manual assessment was conducted by two independent assessors.
Next, we will describe the approach to how our rating agreements were deciphered, how disagreements were resolved, and the key observations made during our assessment of the identified issues.
To quantify our agreement process, we calculated the agreement rate for each project as the percentage of bugs where both evaluators independently agreed on whether the issue qualified as a hot fix.
Across the ten projects we analysed, the average agreement rate was 91\%, indicating a high level of consensus while highlighting areas where subjective judgment played a role.
The agreement rates for individual projects are presented in Table~\ref{tab:hotFixFilter}.

These results reflect variability in how clearly the bugs fit the hot fix criteria, which in turn influenced the rate of agreement.
Apache Flink had the lowest agreement rate at 78.95\%,\ while Apache Jackrabbit had the highest at 100\%.
Disagreements were systematically reviewed through discussions, where both evaluators presented their rationale, referencing the specific criteria outlined in the hot fix definition.
% The primary points of contention revolved around the interpretation of time-criticality.
Common conflict areas included development- related blocker issues such as missing packages, broken builds, and failing CI tests. Although initially excluded as environment problems, they were ultimately classified as hot fixes due to their potential impact on production stability.
Additionally, licensing issues with legal implications, first dismissed as documentation-related, were reclassified as hot fixes given their importance for compliance and operational continuity.
Indeterminate cases without clear consensus were excluded to ensure that only confidently validated hot fixes were retained, preserving the dataset’s integrity.
% \samara{Key observations that emerged during the analysis process: (1) Internal vs. External hotfixes: One evaluator's categorisation method distinguished between internal and external hot fixes based on whether the issue pertained to self-contained program code or depended on external libraries, APIs, or other integrations. (2) Non-critical production vs. Development issues: A recurring theme was the differentiation between issues affecting the production environment versus those impacting development. One evaluator's stricter interpretation often excluded development-related issues unless they directly impeded the production system’s operation.}
Through these detailed evaluations and discussions, we refined our understanding of what constitutes a hot fix, ensuring the benchmark accurately reflects real-world hot fix scenarios.

\begin{table}
  \caption{Inter-rate agreement of manual evaluation. High and consistent across projects confirming robustness.} 
  \centering
  \label{tab:hotFixFilter}
  \begin{tabular}{lrrrrrrrrr}
    \toprule 
    Software  & \#Hot Fixes & \#Hot Fixes & Agreement \\
    & Per Definition & Per Manual & Rate \\
    \midrule
     Flink   & 38 & 34 & 79\%\\
    \midrule    
     Kafka  & 15 & 13 & 87\%\\
    \midrule
     Hadoop & 105 & 95 & 86\%\\
     \midrule
     Solr & 74 & 66 & 92\%\\
     \midrule
     Hbase &  23 & 22 & 96\%\\
     \midrule
    % Cassandra& & & & & &\\
    % \midrule
    %  Fineract& & & & & &\\
    % \midrule
    %  Hive& & & & & &\\
    % \midrule
     Jackrabbit  &  27 & 27 & 100\%\\
    \midrule
    %  Jackrabbit-Oak& & & & & & \\
    % \midrule
     Karaf  &  21 & 20 & 90\%\\
    \midrule
    Calcite & 16 & 13 &  88\% \\
    \midrule
    Nifi  & 67 & 64 & 94\% \\
    \midrule
    Ambari  & 360 & 325 & 92\% \\
     % Spark & & & & \\
     % Groovy & & & & \\
    \bottomrule
    \textbf{Total}  & \textbf{746} & \textbf{679} &\\
    \textbf{Weighted Average} & &  & \textbf{91\%}\\    
    \bottomrule
  \end{tabular}
\end{table}

\begin{table*}
  \caption{Hot Fix Integration. 110 hot fixes were integrated. Most exclusions stemmed from missing tests or non-Java patches.} 
  \centering
  \label{tab:hotFixIntegration}
  \begin{tabular}{lrrrrrrrrr}
    \toprule 
    Software & \multicolumn{6}{c}{\#Cannot be integrated per reason} & \#Integrated\\
     & not java & no tests & build issues & tests passing on buggy & change in test only & other\\
    \midrule
     Flink & 7 & 5 & 0 & 2 & 3 & 2 & 15\\
    \midrule    
     Kafka  & 2 & 4 & 1 & 0 & 0 & 0 & 6\\
    \midrule
     Hadoop & 21 & 27 & 3 & 7 & 16 & 0 & 21\\
     \midrule
     Solr&  7 & 19 & 7 & 8 & 10 & 4 & 12\\
     \midrule
     Hbase & 1 & 12 & 0 & 3 & 1 & 0 & 5\\
     \midrule
    % Cassandra & & \\
    % \midrule
    %  Fineract & & \\
    % \midrule
    %  Hive & & \\
    % \midrule
     Jackrabbit& 0 & 15 & 4 & 1& 2& 1 & 4\\
    \midrule
    %  Jackrabbit-Oak& & & & & & \\
    % \midrule
     Karaf& 2& 15& 0& 0& 1& 1 & 1\\
    \midrule
    Nifi & 18 & 23 & 3 & 1& 6 & 2 & 11\\
    \midrule
    Calcite & 2& 2& 3 & 0 & 3 & 0&3\\
    \midrule
    Ambari & 261 & 19 & 2 & 5 & 6& 0 & 32\\
    \midrule
     % Spark & & & & \\
     % Groovy & & & & \\
    \bottomrule
   \multicolumn{7}{c}{\textbf{Total number of bugs added to hot fix benchmark}}  & \textbf{110}\\
    \bottomrule
  \end{tabular}
\end{table*}

\subsection{Hot Fix Integration}

The HotBugs.jar benchmark builds on the foundation of the widely used Bugs.jar dataset.
Bugs.jar is a benchmark designed to provide reproducible real-world bugs for empirical software engineering research, particularly in automated program repair and fault localization.
In Bugs.jar, each project is forked from its original repository, ensuring that the benchmark preserves the authentic codebase and history.
Each bug is represented as a dedicated branch within the forked repository, where the branch reproduces the bug and its fix.
Within each branch, a hidden folder named .bugs-dot-jar stores important metadata, including the test results report for the failing buggy version, the test results report for the fixed version, and an issue report describing the bug.
This structure standardizes how bugs are stored and reproduced, making it easier for researchers to run experiments and compare results.

To integrate hot fixes into this standardized format, we began by identifying the fix commit associated with each hot fix.
This was typically obtained from the pull request linked to the Jira issue.
Once the fix commit was located, the buggy commit was defined as the commit immediately preceding the fix in the project’s version control history.
Using this pair of commits, we generated a diff file that captures all the changes introduced by the hot fix.
By applying this diff in reverse, we reconstructed the buggy version of the codebase, ensuring that the bug state matched what developers would have encountered when the hot fix was originally applied.
To maintain consistency with Bugs.jar, we adopted the same branch naming convention, but with the addition of a \textit{\_HOTFIX} suffix. This clear labeling allows users to easily locate hot fix branches within the repository and distinguish them from standard bug branches.

Once the buggy version was reconstructed, the next step was to build the project. This process was rarely straightforward, particularly for older bugs, where configurations, dependencies, and build tools had often become outdated or misaligned with modern environments.
Many legacy hot fixes required manual intervention to resolve version mismatches, deprecated plugins, or missing dependencies. 
We iteratively updated project configurations to ensure that the buggy version could be built successfully while maintaining fidelity to the original environment.
These adjustments were carefully documented to ensure they could be replicated and to avoid introducing artificial changes unrelated to the original bug.

After achieving a successful build, we focused on test validation.
For each buggy version, the test suite was executed to confirm that the tests fail as expected, demonstrating the presence of the bug.
If the tests passed on the buggy version, the hot fix was excluded from the benchmark because it could not be reliably reproduced.
Next, the tests were run on the fixed version.
If any tests failed on the fixed version, these were identified as flaky tests and were removed to ensure consistency.
Table ~\ref{tab:hotFixIntegration} summarizes the integration process and details the reasons why certain bugs could not be integrated.
The final output of this process included the failing test results for the buggy version, which were stored in the \textit{.bugs-dot-jar} folder alongside the other metadata.
This rigorous validation ensured that each integrated hot fix provides a clear, reproducible example of a time-critical bug and its resolution.

\section{Research Opportunities}
HotBugs.jar is the first dataset of hot fixes collected.
It opens new opportunities to study the unique nature of time-critical bugs and their remediation practices.
Most directly, it enables the development and benchmarking of automated repair techniques designed for rapid patching under production constraints.
Its rich metadata further supports predictive modeling to identify issues likely to require hot fixes, enhancing triaging and prioritization processes.
Researchers can also use the dataset to study developer behavior and decision-making under time pressure, revealing how emergency fixes differ from standard maintenance practices.
Additionally, HotBugs.jar facilitates the evaluation of regression testing, fault localization, and CI/CD pipelines in realistic, time-sensitive contexts.
Finally, it provides a foundation for examining the reliability and long-term impact of production hot fixes, guiding safer deployment and rollback strategies.
These are only initial directions that begin to reveal the vast research potential in this underexplored domain.

\section{Limitations and Challenges}
While HotBugs.jar was carefully constructed, several threats to validity remain.
Internal validity: Some hot fixes may have been misclassified despite rigorous manual validation. To mitigate this, two independent assessors reviewed each case, achieving 91\% inter-rater agreement. Disagreements were resolved through discussion, and unresolved cases were discarded to minimize false positives.
Construct validity: Our definition of a hot fix relies on Jira metadata (e.g., issue priority and timestamps), which may occasionally be inaccurate or inconsistently maintained, potentially leading to misclassification or omission of true hot fixes.
External validity: The dataset covers ten Apache projects, which, while diverse, may not fully represent all software domains or proprietary development practices.
Reproducibility: Some older bugs required configuration updates to build successfully.
These may introduce minor deviations from the original environments, but we have documented all changes for transparency.
Despite these limitations, the dataset’s structured design, validation rigor, and early community adoption demonstrate its reliability and practical value.

\section{Conclusions}
We introduced HotBugs.jar, the first benchmark focused on hot fixes.
We mined 746 software patches that fit the definition and found 679 genuine hot fixes after manual evaluation which we made available with their metadata in our dataset.
For reasons we detail, we were able to integrate 110 hot fixes that were successfully reproduced using a test suite
Each hot fix was identified through Jira data, manually validated, and packaged with buggy and fixed versions, tests, and metadata.
HotBugs.jar provides a unique resource for research on rapid debugging and automated repair.
Future work includes integrating the dataset into a testing framework to simplify tool evaluation and extension to other programming languages.

\section*{Acknowledgement}
We thank Samara Banday and Rahim Ali for helping assess Jira issues and patches with one of the authors.
This project has obtained ethics approval by UCL Project ID 1031.

\bibliographystyle{ACM-Reference-Format}
\bibliography{refs, related_work}

\end{document}